\documentclass[prl,twocolumn,showpacs,floatfix]{revtex4}
\usepackage{amssymb}

\usepackage{amsmath}

\def\be{\begin{eqnarray} &&}

\def\ee{\end{eqnarray}}

\def\bew{\begin{widetext}}
\def\ew{\end{widetext}}

\begin{document}

\title{Diquark correlations in baryon spectroscopy and holographic QCD}
\author{Hilmar Forkel$^{1}$ and Eberhard Klempt$^{2}$}
\affiliation{$^{1}$Institut f\"{u}r Physik, Humboldt-Universit\"{a}t zu Berlin, D-12489
Berlin, Germany \\
$^{2}$Helmholtz-Institut f\"{u}r Strahlen- und Kernphysik, Universit\"{a}t
Bonn, D-53115 Bonn, Germany}

\begin{abstract}
We introduce an improved mass formula for the nucleon and delta resonances
and show how it emerges from AdS/QCD in a straightforward extension of the
``metric soft wall'' gravity dual. The resulting spectrum depends on just
one adjustable parameter, characterizing confinement-induced IR deformations
of the anti-de Sitter (AdS)\ metric, and on the fraction of ``good'' (i.e.
maximally attractive)\ diquarks in the baryon's quark model wave function.
Despite its simplicity, the predicted spectrum describes the masses of all
48 observed light-quark baryon states and their linear trajectory structure
with unprecedented accuracy.
\end{abstract}

\maketitle


The arguably most prominent and pervasive structure in the known hadron
spectrum consists of (approximately) linear Regge-type trajectories of equal
slopes on which the square masses $M^{2}$ of excited states organize
themselves in Chew-Frautschi plots, i.e. as a function of either the angular
momentum $L$ (or total spin $J$) or the radial excitation level $N$. The
QCD-based understanding of these trajectories and their relation to linear
quark confinement remains one of the pre-eminent challenges of
strong-interaction physics.

In this Letter we propose a mass formula for the trajectories in the
light-quark baryon (i.e. nucleon and delta) sector which improves upon the
recent AdS/QCD \cite{revs} ``metric soft-wall'' (ms) prediction \cite{for07} 
\begin{equation}
M_{N,L}^{\left( \text{ms}\right) 2}=4\lambda ^{2}\left( N+L+\frac{3}{2}%
\right)  \label{mms}
\end{equation}%
and show how it emerges from a straightforward extension of the ms gravity
dual. More specifically, while Eq. (\ref{mms}) works very well in the $%
\Delta $ sector (i.e. all observed $\Delta ^{\ast }$ resonance states lie
within errors on the predicted trajectory \cite{Klempt:2008rq} whose
empirical slope is used to determine the AdS$_{5}$ deformation parameter $%
\lambda =0.52$ GeV), we will show that its predictions for the nucleon
excitation spectrum can be substantially improved by adding the correction%
\begin{equation}
\Delta M_{\kappa _{\text{gd}}}^{2}=-2\left( M_{\Delta }^{2}-M_{N}^{2}\right)
\kappa _{\text{gd}}  \label{dcmf}
\end{equation}%
which solely depends on the resonances' diquark content. The latter enters
through the good\ diquark fraction $\kappa _{\text{gd}}$ which we define as
the fraction of the space-spin-flavor baryon wavefunction in the SU$\left(
2\right) _{s}\otimes $ SU$\left( 3\right) _{f}$ product representation which
contains the most attractive, i.e. the ``good'' ($0^{+}$ color-antitriplet
iso-singlet) diquark. This implies in particular $\kappa _{\text{gd}}$%
\thinspace =\thinspace 0 for all $\Delta $ and spin-3/2 nucleon resonances, $%
\kappa _{\text{gd}}$\thinspace =\thinspace 1/4 for spin-1/2 nucleons in the
70$_{\text{SU}\left( 6\right) }$ representation and $\kappa _{\text{gd}}$%
\thinspace =\thinspace 1/2 for nucleons in the 56$_{\text{SU}\left( 6\right)
}$ multiplet. The mass correction (\ref{dcmf}) renders the long-suggested %
\cite{ida66} and recently reemphasized \cite{sel06,jaf05} importance of
attractive diquark correlations in baryon structure explicit. Interest in
such correlations currently extends to phenomena ranging from exotic hadrons %
\cite{jaf05} to color-superconducting phases of ultradense matter \cite%
{raj01}.

In order to compare Eqs. (\ref{mms}), (\ref{dcmf}) to experimental data, one
needs to assign intrinsic orbital and spin angular momenta $L$ and $S$ to
the observed states. At first glance this seems unfeasible since only total
spin $J$ and parity $P$ are measured. However, the data show regularities
which bring additional order into the spectrum (see Table \ref{masses}). The
three-body dynamics of the quark model implies that the number of expected
states grows dramatically with increasing excitation energy. We will discuss
these states in the non-relativistic harmonic oscillator (h.o.)
approximation. Although the h.o. states may mix, we will show that the
leading $L$ and $S$ configurations can nevertheless be determined. In terms
of the Jacobi coordinates $\rho $ and $\lambda $, the two h.os. exhibit
orbital $\left( l_{\rho },l_{\lambda }\right) $ and radial $\left( n_{\rho
},n_{\lambda }\right) $ excitations with $L$=\thinspace $l_{\rho }$+$%
l_{\lambda }$ and $N$=\thinspace $n_{\rho }$+$n_{\lambda }$, and the energy
levels develop a shell structure which manifests itself experimentally in a
series of resonance regions.

The Particle Data Group lists 22 nucleon and 22 $\Delta $ resonances with
known $J$ and $P$ \cite{Amsler:2008zz}. We use all of them except for the
one-star $\Delta (1750)$ -- which is special and requires a devoted
discussion -- and except for the two-star $\Delta (2000)$ with $J^{P}$%
=\thinspace $5/2^{+}$ for which three mass values are quoted, 1724, 1752,
and 2200\thinspace MeV. The long debated question whether the Roper
resonance $N(1440)$ is the first radial excitation of the nucleon has
recently found an affirmative answer in electro-production experiments \cite%
{Aznauryan:2008pe}. $\Delta (1600)$ with $J^{P}$=\thinspace $3/2^{+}$ is the
analogue state in the $\Delta $ spectrum; we asign $L$,\thinspace $N$=0,1 to
both. We further conjecture $\mathbf{(A)}$ that radial and orbital
excitations cost the same energy, in line with the $N+L$ dependence of Eq.~(%
\ref{mms}).

The two states $N(1535)$ and $N(1520)$ with $J^{P}$=$1/2^{-}$, $3/2^{-}$,
and the three states $N(1650)$,\thinspace $N(1700)$,\thinspace $N(1675)$
with $J^{P}$=$1/2^{-}$,\thinspace $3/2^{-}$,\thinspace $5/2^{-}$ naturally
form a spin doublet with $L$=1,$S$=1/2 and a $L$=1,$S$=3/2 triplet. Symmetry
arguments forbid a symmetric spin wave function for the lowest-mass $\Delta $
states with $L$=1, and indeed only the $L$=1,\thinspace $S$=1/2 spin doublet 
$\Delta (1620)$, $\Delta (1700)$ is observed. $L$=1,\thinspace $S$=3/2 would
form a triplet; symmetry requires excitation of the radial wave function to $%
N$=1. These states are seen as $\Delta (1900)$,\thinspace $\Delta (1940)$,
and $\Delta (1930)$ having $J^{P}$=$1/2^{-}$,\thinspace $3/2^{-}$,\thinspace 
$5/2^{-}$ quantum numbers. Their low mass -- compared to quark model
calculations -- arises from conjecture $\mathbf{(A)}$. Furthermore, we
conjecture $\mathbf{(B)}$ that all negative parity $\Delta $ excitations
with $S$=3/2 have $N$=1.

\begin{table}[tp]
\caption{Masses and quantum numbers of $N$ and $\Delta $ states, together
with the predictions of Eqs. (\ref{mms}) and (\ref{dcmf}). \protect\vspace{%
-8mm}}
\label{masses}
\begin{center}
\renewcommand{\arraystretch}{1.1} {\footnotesize 
\begin{tabular}{ccccccccr}
\hline\hline
$L,N$ & $\mathit{\kappa }_{\text{gd}}$ & \multicolumn{4}{c}{Resonance} &  & 
Pred. &  \\ \hline
0,0 & $\frac{1}{2}$ & $N(940)$ &  &  &  & \hfill input: & \textbf{0.94} & 
\\ 
0,0 & 0 & $\Delta (1232)$ &  &  &  &  & 1.27 &  \\ 
0,1 & $\frac{1}{2}$ & $N(1440)$ &  &  &  &  & 1.40 &  \\ 
1,0 & $\frac{1}{4}$ & $N(1535)$ & $N(1520)$ &  &  &  & 1.53 &  \\ 
1,0 & 0 & $N(1650)$ & $N(1700)$ & $N(1675)$ &  &  & 1.64 &  \\ 
1,0 & 0 & $\Delta (1620)$ & $\Delta (1700)$ &  & \hspace{-1mm}$L,N$=0,1: & $%
\Delta (1600)$ & 1.64 &  \\ 
2,0 & $\frac{1}{2}$ & $N(1720)$ & $N(1680)$ &  & \hspace{-1mm}$L,N$=0,2: & $%
N(1710)$ & 1.72 &  \\ 
2,0 & 0 & $N(1900)$ & $N(1990)$ & $N(2000)$ & $\Delta (1910)$ & $\Delta
(1920)$ & 1.92 &  \\ 
2,0 & 0 & $\Delta (1905)$ & $\Delta (1950)$ & $\Delta (1900)^{\ast }$\hspace{%
-1mm} & $\Delta (1940)^{\ast }$\hspace{-1mm} & $\Delta (1930)^{\ast }$%
\hspace{-1mm} & 1.92 &  \\ 
0,3 & $\frac{1}{2}$ & $N(2100)$ &  &  &  &  & 2.03 &  \\ 
3,0 & $\frac{1}{4}$ & $N(2070)$ & $N(2190)$ & \hspace{-1mm}$L,N$=1,2: & $%
N(2080)$ & $N(2090)$ & 2.12 &  \\ 
3,0 & 0 & $N(2200)$ & $N(2250)$ & $\Delta (2223)$ & $\Delta (2200)$ & 
\hspace{-1mm}$\mid \hspace{-1mm}L,N$=1,2: & 2.20 &  \\[-1.2ex] 
&  &  &  &  &  & \hspace{-2.5mm}\underline{$\mid \hspace{1mm}\Delta (2150)%
\hspace{-1mm}$} &  &  \\[-2.2ex] 
4,0 & $\frac{1}{2}$ & $N(2220)$ &  &  &  &  & 2.27 &  \\ 
4,0 & 0 & $\Delta (2390)$ & $\Delta (2300)$ & $\Delta (2420)$ & \hspace{-1mm}%
$\mid \hspace{-1mm}L,N$=3,1: & $\Delta (2400)$ & 2.43 &  \\[-1.2ex] 
&  &  &  &  & \hspace{-2mm}\underline{$\mid \phantom{\Delta(2300)}$} & 
\hspace{-10mm}\underline{\hspace{10mm}$\Delta (2350)$} &  &  \\[-2.2ex] 
5,0 & $\frac{1}{4}$ & $N(2600)$ &  &  &  &  & 2.57 &  \\ 
6,0 & $\frac{1}{2}$ & $N(2700)$ &  &  &  &  & 2.71 &  \\ 
6,0 & 0 & $\Delta (2950)$ &  &  & \hspace{-1mm}$L,N$=5,1: & $\Delta (2750)$
& 2.84 &  \\ \hline\hline
\end{tabular}
\renewcommand{\arraystretch}{1.0} } \vspace{-4mm}
\end{center}
\par
$^*$: $L,N$=1,1.\phantom{tttttttttttttttttttttttttttttttttttttttt} \vspace{%
-4mm}
\end{table}

In addition, there are three positive-parity nucleon resonances~$N(1900)$%
,\thinspace $N(1990)$,\thinspace $N(2000)$~with quantum num\-bers $J^{P}$%
=\thinspace $3/2^{+}$,\thinspace $5/2^{+}$,\thinspace $7/2^{+}$, four $%
\Delta $ states $\Delta (1910)$, $\Delta (1920)$,\thinspace $\Delta (1905)$%
,\thinspace $\Delta (1950)$ with $J^{P}$=\thinspace $1/2^{+}$,\thinspace $%
3/2^{+}$,\thinspace $5/2^{+}$, $7/2^{+}$, and no other close-by state with
these quantum numbers. It is therefore natural to assign $L$=2 and spin $S$%
=3/2, yielding a $\Delta $ and a nucleon quartet; the latter quartet is
completed by the recently proposed $N(1880)$ with $J^{P}$=\thinspace $%
1/2^{+} $ \cite{Horn:2008}.

The $J^{P}$=\thinspace $1/2^{+}$ $N(2100)$ has no spin-parity partner; we
interpret it as the third excitation in the series $N(940)$, $N(1440)$%
,\thinspace $N(1710)$,\thinspace $N(2100)$. The two states $N(2070)$, $%
N(2190)$ could form a $L$=3,\thinspace $S$=1/2 spin doublet, and the three
states $N(2200)$,\thinspace $N(2190)$,\thinspace $N(2250)$ an incomplete
quartet where the lowest angular momentum state with $J^{P}$\thinspace
=\thinspace $3/2^{-}$ is missing \cite{Bartholomy:2007zz}. The $N(2190)$
assignment is ambiguous. The $J^{P}$=\thinspace $1/2^{-}$ $N(2080)$ must
have $L$=1 and could be the second radial excitation of $N(1535)$. Its spin
doublet partner is then $N(2090)$ as the second $N(1520)$ excitation, its
isospin partner is $\Delta (2150)$. The isospin partner of $N(2190)$ is
missing. First excitations have been suggested: $N(1905)$ with $J^{P}$%
=\thinspace $1/2^{-}$ \cite{Hohler:1979yr,Manley:1992yb} and $N(1960)$ with $%
J^{P}$=\thinspace $3/2^{-}$ \cite{Cutkosky:1980rh,Manley:1992yb,Horn:2008}.
Their predicted mass is 1.82\thinspace GeV.

Due to its low mass, the $J^{P}$=\thinspace $9/2^{-}$ state $N(2250)$ cannot
have $L$\thinspace =\thinspace 5 (apart from a small admixture), rather it
must have $L$\thinspace =\thinspace 3,\thinspace $S$\thinspace =\thinspace
3/2. The three other members with lower $J$ are missing. $\Delta (2223)$
(suggested in \cite{Arndt:2006bf}) and $\Delta (2200)$ with $J^{P}$%
=\thinspace $5/2^{-}$,\thinspace $7/2^{-}$ are natural candidates for the $L$%
=3,\thinspace $S$=1/2 doublet. At a bit higher mass, $\Delta (2400)$ with $%
J^{P}$=\thinspace $9/2^{-}$ is observed. According to our conjecture $%
\mathbf{(B)}$, it has $L$=3,\thinspace $S$=3/2, $N$=1. $\Delta (2350)$ with $%
J^{P}$=\thinspace $5/2^{-}$ could be its spin-multiplet partner while $J^{P}$%
=\thinspace $3/2^{-}$,\thinspace $7/2^{-}$ are missing. Conjecture $\mathbf{%
(B)}$ also entails the $L$\thinspace =\thinspace 5,\thinspace $S$\thinspace
=\thinspace 3/2, $N$\thinspace =\thinspace 1 assignment for the high-mass $%
J^{P}$=\thinspace $13/2^{-}$ resonance $\Delta (2750)$. At 2.4\thinspace
GeV, a nearly complete spin multiplet with $L$\thinspace =\thinspace
4,\thinspace $S$\thinspace =\thinspace 3/2 is known: $\Delta (2390)(7/2^{+})$%
, $\Delta (2300)(9/2^{+})$, $\Delta (2420)(11/2^{+})$. The assignments of $%
N(2700)$ and $\Delta (2950)$, finally, are unambiguous.

Table \ref{masses} reveals the extent of the agreement between the 48
measured mass values and Eqs.~(\ref{mms}), (\ref{dcmf}). In fact, the latter
considerably improve upon any dynamical quark model prediction of the full
mass spectrum. This supports the validity of both Eqs.~(\ref{mms}), (\ref%
{dcmf}) and of our quantum number assignments, and it raises the hope that
not only Eq.~(\ref{mms}) but also the universal correction term (\ref{dcmf})
may have a transparent origin in holographic QCD. In the following we will
show that this is indeed the case. Since the metric soft wall\ \cite{for07}
is so far the only AdS/QCD dual which predicts linear square-mass
trajectories for baryons~(cf. Eq. (\ref{mms})), we take this background as
our starting point. The IR deformed AdS$_{5}$ line element 
\begin{equation}
ds^{2}=g_{MN}dx^{M}dx^{N}=a^{2}\left( z\right) \left( \eta _{\mu \nu
}dx^{\mu }dx^{\nu }-dz^{2}\right)  \label{met}
\end{equation}%
(where $\eta $ is the four-dimensional Minkowski metric and $z$ the
conformal coordinate of the fifth dimension) of the metric soft wall in the
baryon sector is characterized by the warp factor \cite{for07}%
\begin{equation}
a^{\left( \text{ms}\right) }\left( z\right) =\frac{R}{z}\left( 1+\frac{%
\lambda ^{2}z^{2}}{m_{5}^{\left( 0\right) }R}\right)  \label{alt}
\end{equation}%
(where $R$ is the curvature radius, $m_{5}^{\left( 0\right) }$ the Dirac
bulk mode mass and $\lambda $ the scale which governs the IR deformation).
In this geometry the Dirac equation for the string modes $\Psi \left(
x,z\right) $ dual to nucleons reads 
\begin{equation}
\left[ ie_{A}^{M}\Gamma ^{A}\left( \partial _{M}+2a^{\left( \text{ms}\right)
-1}\partial _{M}a^{\left( \text{ms}\right) }\right) -m_{5}^{\left( 0\right) }%
\right] \Psi \left( x,z\right) =0  \label{de}
\end{equation}%
(where $e_{M}^{A}=a\delta _{M}^{A}$ are f\"{u}nfbeins and $\Gamma ^{A}$
Dirac matrices). By iteration, Eq. (\ref{de}) can be cast into the
Sturm-Liouville form%
\begin{equation}
\left[ -\partial _{z}^{2}+V_{\pm }^{\left( \text{ms}\right) }\left( z\right) %
\right] \psi _{\pm }^{\left( \text{ms}\right) }\left( z\right) =M^{\left( 
\text{ms}\right) 2}\psi _{\pm }^{\left( \text{ms}\right) }\left( z\right)
\label{sl}
\end{equation}%
where the modes $\psi _{\pm }\left( z\right) =\left( \lambda ze^{-A}\right)
^{-2}\varphi _{\pm }\left( z\right) $ describe the rescaled chiral
components of the bulk spinor 
\begin{eqnarray}
\Psi \left( x,z\right) &=&\int \frac{d^{4}k}{\left( 2\pi \right) ^{4}}%
e^{-ikx}\times  \notag \\
&&\times \left[ \varphi _{+}^{\left( k\right) }\left( z\right) P_{+}+\varphi
_{-}^{\left( k\right) }\left( z\right) P_{-}\right] \hat{\Psi}^{\left(
4\right) }\left( k\right)
\end{eqnarray}%
($P_{\pm }\equiv \left( 1\pm \gamma ^{5}\right) /2$) and the boundary spinor 
$\hat{\Psi}^{\left( 4\right) }$ solves the $d=4$ Dirac equation $\left(
\gamma ^{a}k_{a}-\left| k\right| \right) \hat{\Psi}^{\left( 4\right) }\left(
k\right) =0$. The metric soft-wall potential is \cite{for07} 
\begin{equation}
V_{m_{5}^{\left( 0\right) },\pm }^{\left( \text{ms}\right) }\left( z\right)
=\pm \partial _{z}\left( a^{\left( \text{ms}\right) }m_{5}^{\left( 0\right)
}\right) +a^{\left( \text{ms}\right) 2}m_{5}^{\left( 0\right) 2}
\label{mpot}
\end{equation}%
and the corresponding eigenfunctions $\psi _{\pm }^{\left( \text{ms}\right)
} $ can be found analytically \cite{for07}. After implementing the boundary
behavior $\varphi _{+}\left( z\right) \overset{z\rightarrow 0}{%
\longrightarrow }z^{\bar{\tau}}$ required by the AdS/CFT dictionary for
baryon interpolators of twist dimension $\bar{\tau}$, which amounts to
replacing $m_{5}^{\left( 0\right) }R\ \rightarrow \bar{\tau}-2=L+1$ \cite%
{bro06}, the eigenvalue spectrum (\ref{mms}) emerges.

As mentioned above, this spectrum agrees very well with data in the $\Delta $
sector but can be substantially improved in the nucleon sector by adding the
correction term (\ref{dcmf}) which depends on the diquark content of the
baryon resonances. At first it might seem impossible to gain holographic
access to this diquark information because diquarks and their operators are
gauge dependent whereas only gauge-invariant operators have well-defined
duals. However, diquark information may enter indirectly through
gauge-invariant baryon interpolating fields whose general form \cite{esp83}%
\begin{equation}
\eta _{t}\left( x\right) =2\left[ \eta _{\text{pd}}\left( x\right) +t\eta _{%
\text{sd}}\left( x\right) \right]  \label{ni}
\end{equation}%
at leading twist (i.e. with the minimal scaling dimension $9/2$) contains a
pseudoscalar diquark in $\eta _{\text{pd}}=\varepsilon _{abc}\left(
u_{a}^{T}Cd_{b}\right) \gamma ^{5}u_{c}$ and a ``good'' scalar diquark in $%
\eta _{\text{sd}}=\varepsilon _{abc}\left( u_{a}^{T}C\gamma ^{5}d_{b}\right)
u_{c}$. (Higher-twist interpolators contain additional covariant
derivatives.)

The interpolators (\ref{ni}) are expected to have enhanced overlap with
nucleon states of the corresponding diquark content and can thus be related
to the $\kappa _{\text{gd}}$ of these states \footnote{%
The analoguous dependence of nucleon correlators on inter\-polators with
different diquark content (and their different\hspace{-0.2mm} couplings%
\hspace{-0.2mm} to\hspace{-0.2mm} instantons)\hspace{-0.2mm} was\hspace{%
-0.2mm} exploited\hspace{-0.2mm} in QCD\hspace{-0.2mm} sum-rule analyses
including small-scale instanton effects \cite{for93}, in instanton vacuum
models \cite{sch94} and on the lattice \cite{deg08}.}$.$ Since $t\rightarrow
\infty $ corresponds to maximal ``good'' diquark content and $t=-1$ (the
``Ioffe current'') to an exclusively ``bad'' diquark content, a simple
approximation to this relation may e.g. be%
\begin{equation}
t\left( \kappa _{\text{gd}}\right) =\frac{1}{2\kappa _{\text{gd}}-1}.
\end{equation}%
The different diquark content of the baryon interpolators $\eta _{t}$ can
affect properties of their dual modes in several ways. For once, the AdS/CFT
dictionary relates the mass $m_{5}$ of a dual string mode to the scaling or
twist dimension of the corresponding interpolator. While all equal-twist
baryon interpolators have the same classical twist dimension $\bar{\tau}=L+3$
and hence correspond to dual modes of equal mass $m_{5}^{\left( 0\right) }\
=\left( L+1\right) /R$ \cite{bro06}, their anomalous dimensions will
generally depend on $t$ and therefore induce $\kappa _{\text{gd}}$ dependent
mass corrections $\Delta m_{5}\left( \kappa _{\text{gd}}\right) .$ Such
corrections could also result from $\kappa _{\text{gd}}$ dependent couplings
of the Dirac modes to condensate-related spin-0 bulk fields. Since at
present no reliable QCD information on nonperturbative anomalous dimensions
exist, we will treat $\Delta m_{5}$ as an adjustable parameter to be
determined in bottom-up fashion from the observed baryon spectrum (as
similarly suggested in Ref. \cite{hon06}).

We are now going to show that such a bulk mass correction $\Delta
m_{5}\left( \kappa _{\text{gd}}\right) $ indeed results in the desired
spectral correction (\ref{dcmf}) and minimally just requires a small
additional IR deformation of the soft wall metric (\ref{met}). To this end,
we first observe that a shifted dual mode mass%
\begin{equation}
m_{5}=m_{5}^{\left( 0\right) }+\Delta m_{5}=\frac{L+\Delta m_{5}R+1}{R}
\end{equation}%
with the corresponding IR deformation%
\begin{equation}
a\left( z\right) =\frac{R}{z}\left( 1+\frac{\lambda ^{2}z^{2}}{L+\Delta
m_{5}R+1}\right)  \label{a}
\end{equation}%
of the warp factor (\ref{alt}) turns the Sturm-Liouville potential (\ref%
{mpot}) into%
\begin{equation}
V_{L,\pm }\left( z\right) =V_{m_{5}^{\left( 0\right) }+\Delta m_{5},\pm
}^{\left( \text{ms}\right) }\left( z\right) =V_{L+\Delta m_{5}R,\pm
}^{\left( \text{ms}\right) }\left( z\right) .
\end{equation}%
Hence the eigenfunctions $\psi _{\pm }$ in this potential can be obtained
from the analytical solutions $\psi _{N,L,\pm }^{\left( \text{ms}\right) }$
of Ref. \cite{for07} as $\psi _{N,L,\kappa _{\text{gd}},\pm }\left( z\right)
=\psi _{N,L+\Delta m_{5}R,\pm }^{\left( \text{ms}\right) }\left( z\right) $
(which are well-defined for $\Delta m_{5}R>-3/2$). Finally, after specifying%
\begin{equation}
\Delta m_{5}\left( \kappa _{\text{gd}}\right) =\frac{\Delta M_{\kappa _{%
\text{gd}}}^{2}}{4\lambda ^{2}R}  \label{dm5}
\end{equation}%
(so that $\Delta m_{5}R\gtrsim -0.7\ $and $m_{5}>0$) the eigenvalue spectrum
(\ref{mms}) indeed turns into%
\begin{equation}
M_{N,L}^{2}=M_{N,L+\Delta m_{5}R}^{\left( \text{ms}\right) 2}=4\lambda
^{2}\left( N+L+\frac{3}{2}\right) +\Delta M_{\kappa _{\text{gd}}}^{2}.
\label{dcbms}
\end{equation}%
The above derivation also sheds new light on the physics behind this
spectrum. Indeed, it shows that the dual modes $\psi _{N,L,\kappa _{\text{gd}%
},\pm }$ corresponding to larger $\kappa _{\text{gd}}$ (with identical $N$, $%
L$) begin to feel the soft wall at smaller $z$ (cf. Eq. (\ref{a})) and
therefore extend less into the fifth dimension. This reflects the additional
attraction and translates into a smaller size of baryon states with larger $%
\kappa _{\text{gd}}$.

A more elaborate alternative for generating the spectrum (\ref{dcbms}) would
take the RG flow of the anomalous dimensions (or additional bulk fields)
into account \footnote{%
We\hspace{-0.2mm} shall\hspace{-0.2mm} tentatively\hspace{-0.2mm} assume%
\hspace{-0.2mm} that\hspace{-0.2mm} the RG mixing\hspace{-0.2mm} among
anomalous\hspace{-0.2mm} dimensions remains\hspace{-0.2mm} small\hspace{%
-0.2mm}\hspace{-0.2mm} in\hspace{-0.2mm} the \hspace{-0.2mm} restricted $\mu 
$\hspace{-0.2mm} (resp.\hspace{-0.2mm} $z$)\hspace{-0.2mm} range\hspace{%
-0.2mm} of\hspace{-0.2mm} interest\hspace{-0.2mm} ($z\lesssim \lambda ^{-1}$)%
\hspace{-0.2mm} and\hspace{-0.2mm} does\hspace{-0.2mm} not\hspace{-0.2mm}
modify\hspace{-0.2mm} the\hspace{-0.2mm} qualitative\hspace{-0.2mm} results.}%
. According to the gauge/gravity correspondence, this renormalization-scale
dependence translates into a $z$ dependent $\Delta m_{5}\left( z\right) $.
Since the above adaptation of the eigenfunctions then ceases to work, one
may instead attempt to generate the mass correction directly in the mode
potential, 
\begin{equation}
V_{L,\pm }\left( z\right) =V_{L,\pm }^{\left( \text{ms}\right) }\left(
z\right) +\Delta M_{\kappa _{\text{gd}}}^{2},  \label{vcorr}
\end{equation}%
while both the warp factor (\ref{alt})\ and the eigenfunctions of the metric
soft wall remain unchanged. Since the $z$ dependent mass correction $\Delta
m_{5,\pm }\left( z\right) \equiv \tilde{m}_{\pm }\left( z\right) $ results
in the modification%
\begin{equation}
\Delta V_{\pm }\left( z\right) =\pm \left( a^{\left( \text{ms}\right) }%
\tilde{m}_{\pm }\right) ^{\prime }+a^{\left( \text{ms}\right) 2}\tilde{m}%
_{\pm }\left( 2m_{5}^{\left( 0\right) }+\tilde{m}_{\pm }\right)  \label{dv}
\end{equation}%
(the prime denotes $\partial _{z}$) of the soft-wall mode potential (\ref%
{mpot}), a $z$ (and $L$) independent shift $\Delta V_{\pm }=\Delta M_{\kappa
_{\text{gd}}}^{2}$ is obtained when the $\tilde{m}_{\pm }\left( z\right) $
solve the nonlinear differential equations 
\begin{eqnarray}
0 &=&\pm \tilde{m}_{\pm }^{\prime }+\left( 2\frac{L+1}{R}a^{\left( \text{ms}%
\right) }\pm \frac{a^{\left( \text{ms}\right) \prime }}{a^{\left( \text{ms}%
\right) }}\right) \tilde{m}_{\pm }  \notag \\
&&+a^{\left( \text{ms}\right) }\tilde{m}_{\pm }^{2}-\frac{\Delta M^{2}\left(
\kappa _{\text{gd}}\right) }{a^{\left( \text{ms}\right) }}.
\end{eqnarray}%
General solutions of these equations indeed exist and, subject to the
boundary condition $\tilde{m}_{\pm }\left( z\right) \overset{z\rightarrow 0}{%
\longrightarrow }0$, read%
\begin{equation}
\tilde{m}_{\pm }\left( z\right) =\frac{4\left( L+1\right) \Delta
m_{5}\lambda ^{2}z^{2}}{L+1+\lambda ^{2}z^{2}}f_{\pm }\left( z\right)
\label{m}
\end{equation}%
with%
\begin{eqnarray}
f_{+}\left( z\right) &=&\frac{1}{2L+3}\frac{\text{ }_{1}F_{1}\left( 1-\Delta
m_{5}R,L+\frac{5}{2},-\lambda ^{2}z^{2}\right) }{_{1}F_{1}\left( -\Delta
m_{5}R,L+\frac{3}{2},-\lambda ^{2}z^{2}\right) }, \\
f_{-}\left( z\right) &=&\frac{1}{2L+1}\frac{_{1}F_{1}\left( 1+\Delta
m_{5}R,-L+\frac{1}{2},\lambda ^{2}z^{2}\right) }{_{1}F_{1}\left( \Delta
m_{5}R,-L-\frac{1}{2},\lambda ^{2}z^{2}\right) },
\end{eqnarray}%
where the constant $\Delta m_{5}$ is given by Eq. (\ref{dm5}) and $%
_{1}F_{1}\left( a,b,x\right) $ are Kummer functions \cite{abr72}. Inserted
into Eq. (\ref{dv}), the solutions (\ref{m}) reproduce Eq. (\ref{vcorr}) and
hence the spectrum (\ref{dcbms}). However, it is not immediately obvious how
the chirality dependence of the mass functions (\ref{m}) can arise directly
from a modified Dirac mass term.

To summarize, we have presented a light-quark baryon mass formula with only
one adjustable parameter (related to the string tension of linear quark
confinement) which reproduces the masses of all 48 observed nucleon and $%
\Delta $ resonances with far better accuracy than e.g. quark model
predictions (although the latter depend on a substantially larger number of
parameters). In addition, our spectrum relates the trajectory slopes to the $%
\Delta $ ground-state mass and reveals a strikingly systematic role of the
good diquark fraction $\kappa _{\text{gd}}$ in baryon spectroscopy.

We have furthermore shown how this spectrum emerges in holographic QCD,
namely by means of string-mode mass corrections in the metric soft wall
background which can be naturally traced to the varying diquark content of
the QCD baryon interpolators. The latter reflects itself e.g. in different
anomalous dimensions and potentially in diquark-content dependent couplings
of the dual modes to condensate-related spin-0 bulk fields. (It would be
interesting to gain independent quantitative insight into these effects,
e.g. by lattice\ calculations of the QCD baryon interpolators' anomalous
dimensions in the infrared.) Finally, our results indicate that baryon sizes
decrease with increasing good-diquark content.

HF acknowledges financial support from the Deutsche Forschungsgemeinschaft
(DFG).

\end{document}